# UV-Photoinduced Defects In Ge-Doped Optical Fibers

F. Messina[1], M. Cannas[1], K. Médjahdi[2], A. Boukenter[2] and Y. Ouerdane[2]

[1]Dipartimento di Scienze Fisiche ed Astronomiche dell'Università, via Archirafi 36, I-90123 Palermo (Italy)
[2]Laboratoire Traitement du Signal et Instrumentation, UMR 5516 CNRS, Université de Saint-Étienne, 10 rue Barrouin, Bât. F, 42000 Saint-Étienne (France)
Corresponding author : fmessina@fisica.unipa.it

*Abstract* – We investigated the effect of continuous-wave (cw) UV laser radiation on single-mode Ge-doped $H_2$-loaded optical fibers. An innovative technique was developed to measure the optical absorption (OA) induced in the samples by irradiation, and to study its dependence from laser fluence. The combined use of the electron spin resonance (ESR) technique allowed the structural identification of several radiation-induced point defects, among which the Ge(1) ($GeO_4^-$) is found to be responsible of induced OA in the investigated spectral region.

## I - Introduction

Ge-doped glass is a material that has attracted research interest for many years due to its technological importance in both passive and active integrated devices such as optical fibers, Bragg gratings and second harmonic generation [1-2]. One of the peculiar properties leading to these applications, is the photosensitivity of Ge-doped glass under UV exposure [1]. This phenomenon is accompanied, through a process of point defects transformation, by a variation of the OA spectrum [3-6]. Nevertheless, the identification of defects responsible of the OA in irradiated Ge-doped glass is still debated and the photo-induced mechanisms involved in photosensitivity are not yet completely understood.

A convenient experimental approach to investigate point defects embedded in the glass network consists in correlating OA-photoluminescence (PL) and ESR results. In particular, the latter are most suitable for the structural identification of paramagnetic centers [7]. In this work, we report experimental results concerning the effect of UV laser irradiation in single-mode $H_2$-loaded Ge-doped optical fibers. Scientific and applicative interest in $H_2$-loaded fibers is motivated by the circumstance that $H_2$-loading is known to enhance photosensitivity [5, 8]. Comparison with the effects observed in not loaded fibers will be dealt with elsewhere. Our purpose is to contribute to a deeper clarification of the photo-induced processes occurring upon UV irradiation in these materials..

## II – Materials And Methods

Measurements were carried out in commercial single-mode (Corning SMF-28) Ge-doped optical fibers with 4% Ge content in the core. The fiber samples were hydrogen ($H_2$)-loaded in a vessel under a $H_2$ pressure of 100 atm for three weeks at room temperature. Irradiations were carried out at room temperature with a cw frequency-doubled $Ar^+$ laser operating at 244 nm.

To overcome the difficulty of OA measurements in single-mode optical fibers, an innovative technique was used (see Fig. 1 and Ref. 9 for more details). Briefly, the 400-nm photoluminescence (PL), ascribed to the twofold-coordinated Ge, is used as an *in situ* probe source, which is generated during transversal UV exposure of the uncoated fiber [10]. The fiber is translated during irradiation, so that the PL is transmitted along the already-irradiated part of the fiber. Hence, the PL signal is partly absorbed within the irradiated length and its detected amplitude reduction allows measuring the induced OA in the 360-440 nm wavelength range. The PL was detected by a Chromex 500 is/sm spectrometer equipped with an Andor cooled charge-coupled-device camera. Laser fluence is determined by the fiber translation speed, and was varied from 0.1Jcm$^{-2}$ to 5Jcm$^{-2}$. The typical exposure time of a given portion of fiber to laser radiation is in the ms range. ESR measurements were performed by a Bruker EMX spectrometer working at 9.7GHz. Signal was detected on irradiated fiber samples a few meters long, after cutting them in portions of ~0.5cm.

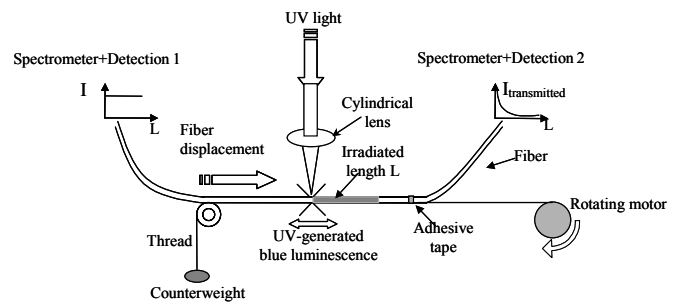

Fig. 1: Experimental set-up used to measure the UV-induced absorption. The beam size on the fiber is 60 μm longitudinally and 800 μm transversally to the fiber axis.

Induced paramagnetic defects were revealed using a microwave power P=1.6mW, and a 100 kHz modulation field of peak-to-peak amplitude $B_m$=0.1m, these values being properly chosen to avoid saturation and distortion of associated ESR signals. The concentration of paramagnetic centers was calculated by comparing the double-integrated ESR spectra with that of Si-E'($\equiv$Si•) centers in a reference sample, where their absolute density was determined with accuracy of ±15% by spin-echo decay measurements [11-12].

## III – RESULTS AND DISCUSSION

In Fig. 2(a) is shown the evolution of the emission band centered around 3.1 eV transmitted within the irradiated part of a fiber sample moving during laser-irradiation. From the spectra we extract the decay curves of the luminescence intensity at a given energy as a function of the irradiated length of the fiber; examples of this procedure are shown in the inset. Hence, the absorption coefficient at each energy is calculated from the decay curves by a linear best-fit. In Fig. 2(b) the induced absorption spectra, calculated in this way are reported for fiber samples irradiated with three different fluences. As apparent from data, the induced OA profile in the investigated spectral range (limited by the width of the 3.1eV emission band) is the low energy tail of an OA band centered at higher energies.

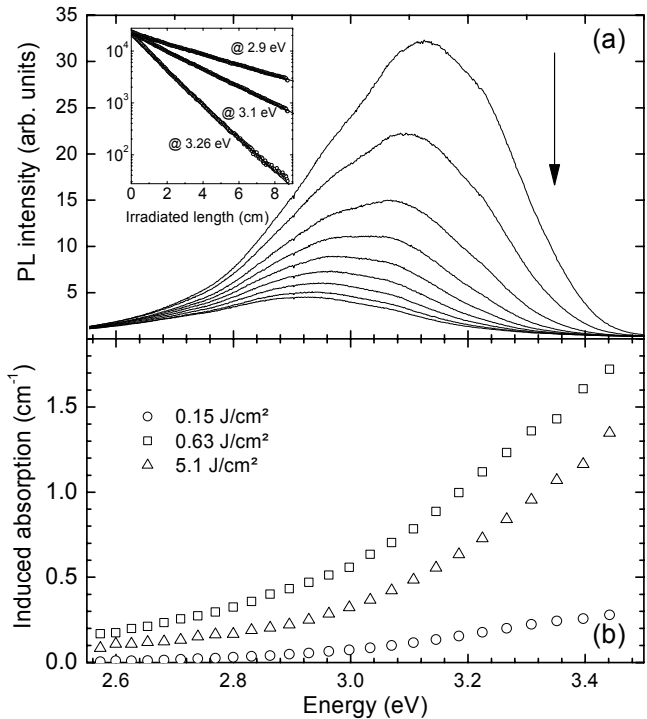

Fig.2: (a) Evolution of the PL spectrum peaked at around 3.1 eV transmitted within the irradiated part of the fiber during irradiation and concurrent movement. The arrow indicates the direction of growing time. The inset depicts the evolution of the PL intensity, for a given energy, extracted from the spectra. From fitting this data we extract the absorption profiles of Fig. 2(b)

Induced OA is a manifestation of conversion processes triggered by UV radiation which generate Ge-related point defects absorbing in this region from precursors.

The dependence from irradiation dose is summarized in Fig. 3, where induced absorption coefficient at 3.1eV, $\Delta\alpha$(3.1eV), is plotted against laser fluence: on increasing fluence up to ~1 J/cm² absorption increases up to the maximum value of ~0.6 cm$^{-1}$, after which it remains invariant. This finding means that the generation process of the absorbing defects saturates at the ~1 J/cm² fluence level, either by exhaustion of precursors or by reaching a dynamical equilibrium. ESR measurements are particularly suitable to find out which centers are responsible of the induced absorption, since this technique permits the unambiguous identification of paramagnetic defects generated by UV exposure. Fig. 4 shows the ESR spectra detected in fibers after irradiation at different fluences. From comparison with literature data, the observed lineshape in the 345-350mT region is considered to derive from the superposition of two Ge-related signals falling in this region, namely Ge(1) and E'-Ge [13-14]. Ge(1) is an electron trapped on a fourfold coordinated Ge ($GeO_4^-$), while E'(Ge) is an electron localized on a threefold Ge ($\equiv$Ge•). The former is mainly responsible of the positive peak, whereas the negative structure is mostly ascribable to the latter. Finally, the hyperfine doublet with 11.8mT isotropic splitting typical of H(II) centers (=Ge•-H) is detected and shown in the inset [15]. Ge(1) and E'(Ge) are very commonly radiation-induced paramagnetic defects in Ge-doped silica, whereas H(II) centers are usually observed only in UV-irradiated $H_2$-loaded Ge-doped samples and fused quartz [3-8,13-17].

A deconvolution procedure was performed to estimate separately the concentration of the different paramagnetic centers which contribute to the ESR signal: to this aim, ESR spectra over the 345-350mT region were expressed as a linear combination of Ge(1) and E'(Ge) lineshapes.

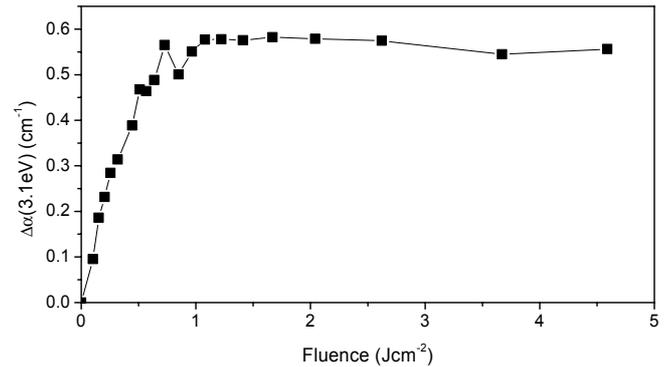

Fig.3: Radiation-induced absorption coefficient at 3.1eV as a function of laser fluence

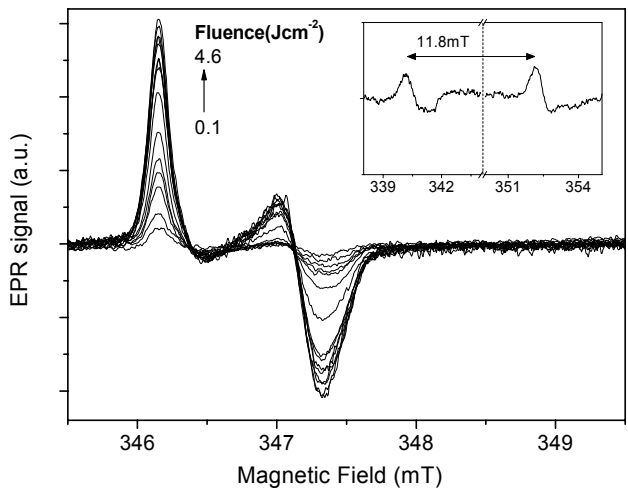

Fig.4: Composite ESR spectra of irradiated optical fibers observed in the 345-350mT spectral region. In the inset: H(II) centers hyperfine doublet, observed scanning a wider magnetic field region (break in the graph extends from 344mT to 350mT).

The two lineshapes used as basis functions are of experimental origin, measured in suitable reference samples where it was possible to isolate them (i.e. Ge(1) in γ-irradiated sol-gel silica samples and E'-Ge in preforms of the same composition of the fibers used in the present study) [17]. The concentration of H(II) centers was estimated by direct numerical double integration of the hyperfine doublet. Results of these calculation are shown in Fig. 5, where the concentration of the three species are plotted against fluence.

We see that the concentrations of Ge(1) and E'(Ge) grow with fluence up to $(1.5\pm0.2)\cdot10^{17}cm^{-3}$ and $(3.0\pm0.4)\cdot10^{17}cm^{-3}$ respectively, whereas H(II) centers have already reached their maximum concentration of $(1.9\pm0.3)\cdot10^{17}cm^{-3}$ at the lowest fluence ($0.1Jcm^{-2}$), from which they slowly decrease down to $(1.4\pm0.2)\cdot10^{17}cm^{-3}$ measured at 4.6 $Jcm^{-2}$. If one of the induced paramagnetic centers is responsible of the induced OA profile in the investigated region, then its concentration should depend from fluence in a similar way to the absorption coefficient. Hence, present data allow to exclude that H(II) centers are responsible of measured OA since they do not grow with fluence in the $0.1Jcm^{-2} \div 4.6Jcm^{-2}$, so showing a strikingly different behaviour from the absorption coefficient (compare with Fig. 3).

To extend this approach to the other two observed paramagnetic species, in Fig. 6 we plot the absorption coefficient at 3.1eV against the concentration of (a) E'(Ge) and (b) Ge(1). The plot shows that only Ge(1) concentration is linearly correlated with OA at 3.1eV within experimental error. At variance, datapoints in the correlation plot for E'(Ge) do not show a linear tendency; indeed, from comparison of Fig. 5 with Fig. 3, the concentration of E'(Ge) continues to grow after $1.0Jcm^{-2}$, where induced absorption has already reached its maximum.

From this findings we infer that Ge(1) is the point defect generated by laser exposure responsible of the tail-shaped trasmission loss in the fiber. This conclusion is consistent with previous literature studies, in which E'(Ge) was suggested to absorb in a band peaked at 6.4eV with $0.8eV \div 1.2eV$ FMWH, whereas the observed tail-shaped profile in the investigated region is consistent with literature parameters of the Ge(1) absorption band, suggested to peak at 4.5eV with $0.9eV \div 1.3eV$ FWMH [6,8,14,17-19]. Now we briefly address the issue of the generation mechanism of Ge(1) center. As known from ESR studies, Ge(1) is induced by trapping of an electron $e^-$ on a fourfold coordinated (GeO$_4$) Ge precursor [4,6,8,14]. Hence, the generation of Ge(1) requires ionization of an electron donor which makes available the charge to be trapped on GeO$_4$.

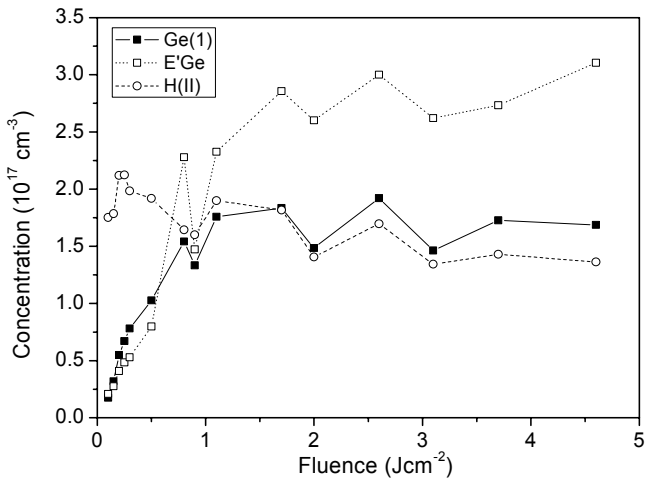

Fig.5: Concentration of the paramagnetic species Ge(1), E'(Ge) and H(II) as a function of fluence.

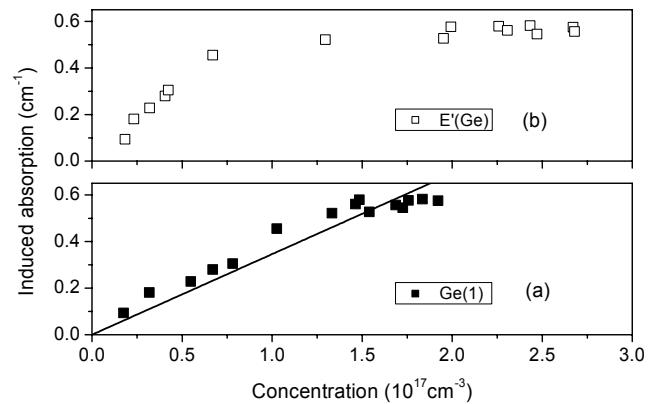

Fig.6: Correlation plot between (a) Ge(1) concentration and induced absorption or between (b) E'(Ge) concentration and induced absorption.

In the present case, the observation of E'(Ge) may give a clue on the e⁻ production process, since E'(Ge) can arise from ionization of a neutral oxygen vacancy on Ge [3]. Though, this hypothesis contrasts with the fact that E'(Ge) and Ge(1) have a different dependence from fluence, suggesting to look for alternative photochemical mechanisms.

An alternative e⁻ source is the twofold coordinated Ge, which is ionized by UV in Ge(2) center (=Ge•) so making free an electron [6,8]. This possibility is suggested by a further experimental result (Fig. 7), i.e. the bleaching of the 3.1 eV PL observed during laser exposure performed *without* moving the fiber which is being irradiated. In this case, the bleaching is obviously not a consequence of induced OA; at variance, it demonstrates the occurrence of laser-induced conversion of the twofold coordinated Ge in other centers, consistently with the proposed Ge(2) model. It is worth to note that the absence of the typical Ge(2) signal in ESR spectra is expected since this paramagnetic species is known to be annealed by $H_2$ even at room temperature, so preventing its observation in our experiment [8]. We also note that annealing of Ge(2) with subsequent e⁻-re-trapping is a possible generation channel for the observed H(II) centers [8].

Additional measurements are required to fully clarify the photochemical mechanism responsible of induced OA.

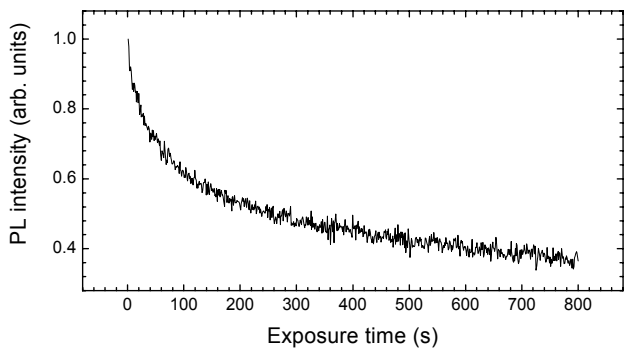

Fig. 7: Evolution of the intensity of the 3.1 eV PL during transversal irradiation of the fiber without simultaneous translation. The time scale of observation here is much longer than the typical irradiation time of Fig. 2.

## IV – CONCLUSIONS

We investigated the effects induced in $H_2$-loaded SMF-28 fibers by cw UV irradiation. A novel measurement technique was developed to perform *in situ* optical measurements in the fiber core. We found a tail-shaped OA profile induced by laser radiation in the 2.5eV÷3.5eV spectral range and studied its dependence from laser fluence, finding a saturation at about 1Jcm⁻². ESR measurements on irradiated specimens reveal the generation of several Ge-related paramagnetic centers, among which only the Ge(1) center ($GeO_4^-$) is well correlated to the measured OA. Based on this finding and on comparison of the OA profile with previously reported results, we infer that the laser-induced absorption in the investigated spectral range is associated with the generation of Ge(1) centers. In addition, PL results show that the twofold-coordinated Ge and the Ge atom coordinated to four O-Si next-nearest-neighbor atoms are involved in the generation of Ge(1) center: the former releases an electron, which is trapped by the latter on the Ge atom, to form the Ge(1) center.

ACKNOWLEDGMENTS

The authors would like to thank S. Agnello, R. Boscaino, G. Buscarino and F.M. Gelardi for useful discussions and G. Lapis and G. Napoli for technical assistance. This work is part of a national project (PRIN2002) supported by the Italian Ministry of University Research and Technology.